\documentclass{article}

\usepackage{PRIMEarxiv}

\usepackage[utf8]{inputenc} 
\usepackage[T1]{fontenc}    
\usepackage{hyperref}       
\usepackage{url}            
\usepackage{booktabs}       
\usepackage{amsfonts}       
\usepackage{nicefrac}       
\usepackage{pifont} 
\usepackage{array} 
\usepackage{lscape}
\usepackage{hyperref}
\usepackage{microtype}      
\usepackage{lipsum}
\usepackage{fancyhdr}       
\usepackage{graphicx}       
\graphicspath{{media/}}     

\title{A Survey on Blockchain-based Supply Chain Finance with Progress and Future directions
}

\author{
  Zhengdong Luo \\
  Xinjiang Technical Institute of Physics and Chemistry, Chinese Academy of Sciences, Urumqi 830011, China. \\
  University of Chinese Academy of Sciences, Beijing 100049, China. \\
  \texttt{luozhengdong21@mails.ucas.edu.cn} \\
}

\begin{document}
\maketitle

\begin{abstract}
  Supply Chain Finance is very important for supply chain competition, which is an important tool to activate the capital flow in the supply chain. Supply Chain Finance-related research can support multiple applications and services, such as providing accounts receivable financing, enhancing risk management, and optimizing supply chain management. For more than a decade, the development of Blockchain has attracted widely attention in various fields, especially in finance. With the characteristics of data tamper-proof, forgery-proof, cryptography, consensus verification, and decentralization, Blockchain fits well with the realistic needs of Supply Chain Finance, which requires data integrity, authenticity, privacy, and information sharing. Therefore, it is time to summarize the applications of Blockchain technology in the field of Supply Chain Finance. What Blockchain technology brings to Supply Chain Finance is not only to alleviate the problems of information asymmetry, credit disassembly, and financing cost, but also to improve Supply Chain Finance operations through smart contracts to intelligent Supply Chain Finance and in combination with other technologies, such as artificial intelligence, cloud computing, and data mining, jointly. So there has been some work in Blockchain-based Supply Chain Finance research for different Supply Chain Finance oriented applications, but most of these work are at the management level to propose conceptual frameworks or simply use Blockchain without exploiting its deep applications. Moreover, there are few systematic reviews providing a comprehensive summary of current work in the area of Blockchain-based Supply Chain Finance. In this paper, we provide an overview of Supply Chain Finance and Blockchain technology, including brief knowledge and respective main applications. Furthermore, we summarize the current hot applications of Blockchain-based Supply Chain Finance and give some future directions with a forward-looking orientation in pursuit. This is the first comprehensive survey that targets the study of Blockchain technology for the Supply Chain Finance from computer science perspective, and also offers a collection of research studies and technologies to benefit researchers and practitioners working in different Supply Chain Finance and Blockchain fields.
\end{abstract}


\section{Introduction}\label{title:Introduction}
Supply Chain Finance has a profound impact on industrial upgrading, capital optimization and cost reduction\cite{Yan-2016-SupplyChainFinance}. Financing difficulties have become an important issue for global trade. According to the data released by WTO in August 2021\footnote{https://data.wto.org/(accessed 23 August 2021)}, the global trade finance gap increased by 15\% to \$1.7 trillion in 2020 compared to the investigation two years ago. As indicated by the Asian Development Bank (ADB)\footnote{https://baijiahao.baidu.com/s?id=1713396224663801178\&wfr=spider\&for=pc. Last accessed: 2022-04-01}, the acceptance rate of Small and Medium Enterprises(SMEs) financing was 23\% and the rejection rate was 40\% of the banked trade finance applications, which was more serious compared with the 54\% acceptance rate and 42\% rejection rate of large enterprises. Of course the economic downturn and the impact of the COVID-19 are important causes of financing difficulties, but at the same time another reason is related to the global level of development on Supply Chain Financing. Supply Chain Finance is an important tool to help supply chain to finance, which can alleviate the problems of long accounting period, insufficient funds to expand business, products occupying funds, etc. Supply Chain Finance plays an important role in co-creating value by adjusting capital flow, optimizing capital flow management, and mitigating financing risks\cite{Hofmann-2005-SupplyChainFinance, Xu-2018-SupplyChainFinance}. On the other hand, Blockchain technology as a new technology is gradually receiving attention in various fields, such as healthcare\cite{holbl2018systematic, zubaydi-2019-healthcare, agbo-2019-healthcare}, smart city\cite{esposito-2021-smartCity, majeed-2021-samrtCity, singh-2021-smartCity}, and Supply Chain Finance\cite{omran-2017-BcSCF, li-2020-BcSCF, yao-2020-BcSCF, wang-2021-BcSCF}. Blockchain is considered as a next-generation information technology that redefines the world\cite{lu-2019-Blockchain}, so it is the duty and obligation of Blockchain technology researchers to use this advanced technology to help Supply Chain Finance operations and management become more intelligent and convenient. For these reasons, the research related to Blockchain-based Supply Chain Finance has been a very important and promising field.

The research on Blockchain technology and Blockchain-based Supply Chain Finance started relatively late and has only been developed in the last decade and more, while as far as we can find the earliest paper on Blockchain-based Supply Chain Finance was published in 2018\cite{chaoyong-2018-BcSCF}. Subsequently, more and more research work has been done, and the depth of research has gone from simple application to exploring the application of innovative Blockchain attributes. Initially, Blockchain-based Supply Chain Finance was only a simple application of Blockchain to more management-level skills. For example, Supply Chain Finance information symmetry sharing\cite{zou-2019-BcSCF}, Supply Chain Finance financing model improvement\cite{yang-2021-BcSCF}, and supply chain financial service platform construction\cite{li-2021-BcSCF}. Many of them present conceptual models and do not have experimental validation or practical testing. This makes the work lack some degree of credibility. Later, scholars' research gradually gets deeper rather than just applying, and start to explore the combination of the characteristics of Blockchain and the business needs of Supply Chain Finance, and apply Blockchain technology to Supply Chain Finance from the computer science level, such as the improvement of Supply Chain Finance data storage method\cite{jiang-2021-BcSCF}, the study of Supply Chain Finance risk management model\cite{liu-2021-BcSCF-risk}, and smart contract to help automate Supply Chain Finance\cite{zhang-2021-BcSCF-smartContract}. In addition, to the best of our knowledge, there is no publicly available Supply Chain Finance dataset online, let alone a Blockchain-based Supply Chain Finance dataset.

Considering that Blockchain technology research is inevitably getting deeper and deeper, we propose a vision of Blockchain computing, which aims to combine different computer technologies with Blockchain applications to different domains, instead of simply regarding Blockchain as a distributed storage database, the goal of Blockchain computing is to expand the storage database into an intelligent computational database. To the best of our knowledge, Professor Raj Jai et al. first proposed the concept of extended Blockchain\footnote{https://www.cse.wustl.edu/\textasciitilde jain/talks/pbc\_iics.htm. Last accessed: 2022-04-01}\footnote{https://www.cse.wustl.edu/\textasciitilde jain/talks/pbc\_ibf.htm. Last accessed: 2022-04-01}, which emphasizes the extension of Blockchain with artificial intelligence for risk management and decision making. However, they do not give a clear definition of Blockchain computing for this purpose. In a broad sense, what we consider as Blockchain computing is not limited to the combination of artificial intelligence, but also data mining, big data, etc. The application field of Blockchain computing is also not just risk management and decision making. We present for the first time here through one future research hot-spot of Blockchain-based Supply Chain Finance. Besides that, there are many  open questions to answer. For example, what is the relationship between supply chain and Supply Chain Finance? What exactly is Blockchain technology? What are the representative applications of Blockchain technology in Supply Chain Finance? What is the method of classifying these applications employed? What are the potential interesting future directions on Blockchain-based Supply Chain Finance?

To answer these questions, we introduce the theory, current research status and future research on Blockchain-based Supply Chain Finance from three modules: Supply Chain Finance, Blockchain technology, and Blockchain+Supply Chain Finance. Some related surveys have been done. For example, For example, \cite{li-2019-BcSCF-survey} gave a simple survey on Blockchain-based Supply Chain Finance risk control. However, it explores the Loan-to-Value rate, financing schemes, and financial derivatives from a management perspective. The survey of the current state of research and future directions presented by\cite{gelsomino-2016-SCF-survey, xu-2018-SCF-survey} was only focused on Supply Chain Finance. As to the best of our knowledge, we are the first survey of Blockchain-based Supply Chain Finance, not specifically in the direction of risk control, and not just an review of Supply Chain Finance, but an overview of Blockchain technology applied to Supply Chain Finance. Also, the current survey of Blockchain-based Supply Chain Finance is indeed rare. This survey aims to provide a comprehensive summary of the current Blockchain-based Supply Chain Finance applications to identify the issues to be addressed and to point out the future directions. It seeks to consider the relationship between Blockchain and Supply Chain Finance from a computer science perspective, to establish a link between Blockchain technology and Supply Chain Finance-related fields, and to provide a good reference for the development of Blockchain technology and Supply Chain Finance-related applications in various fields. Our survey pursues a prospective exploration rather than a comprehensive summary. To this end, about 189 studies are shortlisted and classified in this survey. 

This survey is organized as follows: Section {\ref{title:SCF}} introduces the overview of the basic knowledge and research direction on Supply Chain Finance. Section {\ref{title:blockchain}} introduces the Blockchain knowledge, Blockchain research directions and Blockchain applications. Section {\ref{title:Blockchain-SCF}} and {\ref{title:future}} introduce our literature research method briefly, the progress summary of representative applications and future research on Blockchain-based Supply Chain Finance. We finally concludes the article in Section {\ref{title:conclusions}}.

\section{Supply Chain Finance}\label{title:SCF}
\subsection{Supply Chain Finance overview.} 
With the expanding trend of trade globalization, the supply chain mechanism has become more and more complex, the upstream and downstream enterprises in the supply chain are widely under the pressure of capital flow. The core enterprises, with the dominant control over the status of supply chain business, have caused delays to the billing periods of upstream and downstream enterprises in the consideration of maximum financial returns. The core enterprises are normally large enterprises with favorable reputation for financing, while the financing problem faced by Small and Medium Enterprises in the supply chain is always a big trouble. At the same time, the traditional inter-company competition is transforming into a market competition among supply chains. In addition to the global economic slowdown, the research work on Supply Chain Finance, which aims to improve the financing difficulties of Small and Medium Enterprises and enhance the viability of the supply chain, has become a real-time hot theme, which has a positive effect on improving the efficiency of capital flow operations and reducing the overall operating costs of the supply chain.

Supply Chain Finance has unique value in financing in several aspects. Firstly, Supply Chain Finance can reduce financing risks. Supply Chain Finance provides financing services for upstream and downstream Small and Medium Enterprises with the backing of the credit and transactions of core enterprises in the whole supply chain, which reduces the financing risks of companies. Secondly, Supply Chain Finance can reduce financing cost by solving the problem of financing information asymmetry and optimizing the information evaluation cost before financing. Finally, Supply Chain Finance can also improve the efficiency of financing. Through the optimization of capital in the whole supply chain, it can improve the efficiency of capital use, accelerate the speed of financing and ease the demand of Small and Medium Enterprises for financing timeliness.

\subsection{What is Supply Chain Finance?} 
Supply Chain Finance, which the well-know history of research could be backed to the 1970s, was mentioned by Budin and Eapen to study the relationship between capital flow and trade credit and inventory\cite{budin-1970-SCF}, and was first defined as the integration of financial flows into the physical supply chain to improve supply chain management in the year of 2002\cite{stemmler-2002-SCF}. Subsequently, there has been various versions of the definition of Supply Chain Finance. For example, Gomm defined Supply Chain Finance as optimizing inter-firm financing by reducing costs and accelerating the capital flow\cite{gomm-2010-SCF}, while Wuttke et al. described the goal of supply chain as aligning the flow of funds with the flow of products and information within the supply chain to improve the management of financial flows from a supply chain perspective\cite{wuttke-2013-SCF}. In this paper, we adopt the definition of Supply Chain Finance given by Professor Song Hua, which is defined as a comprehensive financial product and service provided to the upstream and downstream enterprises in the supply chain based on the core customer, with the real trade background, using self-reimbursement trade finance, and closing the capital flow or controlling the physical rights through professional means such as pledge registration of accounts receivable and third-party supervision\cite{Songhua-2015-SCF}. Particularly, Supply Chain Finance is an attractive tool providing solutions to capital management and improving whole supply chain operations. its core idea is to help the upstream and downstream enterprises in the supply chain to finance, with the help of supply chain structure, comprehensive supply chain information, relying on its prepaid accounts, inventory, accounts receivable and others to pledge guarantee for financing, and self-tasting financing with the income from future supply chain transactions as the source of repayment\cite{qu-2018-SCF-risk}. In the Supply Chain Finance framework, there are two major parts: supplier financing and distributor financing. These two parts contain four main forms of Supply Chain Finance specific programs: accounts receivable financing, inventory financing, prepayment financing, and strategic relationship financing\cite{information&Communication-2018-BcSCF, Songhua-2015-SCF}.

\subsection{Research direction of Supply Chain Finance.} 
Here we divide the academic research work on Supply Chain Finance into two categories, One is the perspective of management science, which mainly entails studying Supply Chain Finance from financial attributes and supply chain attributes. The other is the computer science perspective, which uses advanced computer technology to optimize Supply Chain Finance operations. Both of these two categories of research aim at solving the financing problems of supply chain enterprises, facilitating financing as much as possible under the premise of controlling risks, and at the same time ensuring that the financing money is repaid safely. In the management perspective, some researchers had studied the basic theory of Supply Chain Finance in terms of financial attributes\cite{martin2019towards, guo2014research, kouvelis2012financing, tunca2018buyer}and supply chain attributes\cite{chen2019role, yaqin2016coordination, zhang2014supply}, with the former considering the provision of mutually beneficial financing solutions from the standpoint of financial institutions, and the latter emphasizing Supply Chain Finance to improve the supply chain ecology to enhance the competitiveness of the whole supply chain. Another part of scholars focused on the study of financing models of Supply Chain Finance and the impact of financing services on supply chain performance. such as Kouvelis et al.\cite{kouvelis2012financing} and Tunca et al.\cite{tunca2018buyer} to study the impact of different financing channel choices on supply chain performance, while Dou YQ et al.\cite{yaqin2016coordination} and Zhang Q et al.\cite{zhang2014supply} tried to coordinate the supply chain performance by introducing contract to achieve the goal of improving the overall efficiency of the supply chain and "win-win" for all participants. Another major category is the application of advanced computer technology to Supply Chain Finance, such as big data\cite{chen2019iot, 2019Big}, cloud computing\cite{hemanth2017ahp, singh2015data}, artificial intelligence\cite{hu2020statistical, zhu2019forecasting}, and Blockchain\cite{du2020supply, wang2021research}. For example, Chen R Y\cite{chen2019iot} analyzed the risk performance of Supply Chain Finance with big data. Guggilla Hemanth et al.\cite{hemanth2017ahp} researched and explored cloud computing technology to alleviate communication challenges among supply chain participants in order to improve collaboration and communication. Hu Z\cite{hu2020statistical} considered that the knowledge of enterprise background and supply chain data could reduce the degree of information asymmetry and facilitate the selection of Supply Chain Finance models. A typical research example is Du M et al.\cite{du2020supply} using Blockchain for Supply Chain Finance innovation. The authors redesigned the process and business model of warehouse receipt pledge financing and accounts receivable factoring in Supply Chain Finance and proposed a Blockchain-based supply chain service platform structure, while by improved Paillier algorithm and smart contract to design a Blockchain homomorphic encryption scheme to meet the demand of sensitive data privacy protection in Supply Chain Finance scenario. There are also a lot of research work on Blockchain technology being applied to the field of Supply Chain Finance, which we will expand in detail later.

\section{Blockchain Technology} \label{title:blockchain}
\subsection{Blockchain Overview.} 
As this survey is not dedicated to Blockchain knowledge, what is mentioned here is briefly and not explored in depth, aiming to give a broad understanding of Blockchain related points.The mysterious author Satoshi Nakamoto published a paper entitled "Bitcoin: a peer-to-peer Electronic Cash System", marking the birth of Bitcoin\cite{nakamoto2008bitcoin} in 2008. It is currently the first representative application of Blockchain technology with the most influence in the financial field\cite{cheah2015speculative}. The point-to-point model proposed by Satoshi Nakamoto, which has the characteristics of no third-party financial institutions, decentralization, and de-trusted, caused a sensation in this economic field and even others. With the exciting Bitcoin, there is a spirit of discovery to study its underlying technology, namely blockchain technology. Blockchain is rapidly being used in finance, supply chain, healthcare, etc. The research on the application of blockchain technology in the financial industry has also become the most sought after hot spot. Sonal Trivedi et al.\cite{trivedi2021systematic} and Mohamad Osmani et al.\cite{osmani2020blockchain} provided an overview of the progress and trends in the application of blockchain to the financial industry. Aslo, the supply chain is a series of links in the process from the starting supplier/raw material to the end consumer/product, including manufacturing, transmission, storage, distribution, etc. The study of supply chain management technology has become the focus of attention of many researchers\cite{wang2018understanding, shakhbulatov2020blockchain, lim2021literature}. While Blockchain research has been expanding its applications into non-financial areas such as healthcare which is relatively new but growing rapidly. The value sharing of healthcare data is very beneficial to the identification of medical patterns and the improvement of medical technology, but the privacy conservation of healthcare data is an impediment to this. The utility of Blockchain with distributed storage, encryption algorithms and access control has brought light to solve the complex challenges of healthcare data. In recent years, many surveys have systematically summarized the application of Blockchain technology in the medical field\cite{holbl2018systematic, zubaydi2019review, 2019Big, agbo2019blockchain, mcghin2019blockchain}. Therefore, it is essential to investigate the nature of blockchain technology and its progresses. This lays some sort of foundation for our survey of blockchain-based Supply Chain Finance.

\subsection{What is Blockchain?} The Blockchain is originally defined as a data structure that stores historical data of Bitcoin transactions\cite{nakamoto2008bitcoin}. The transaction data for a period of time is packaged to form a new block, and is linked to the previous block according to a certain rules, and many blocks are linked in chronological order to form chain structure, which is vividly called as Blockchain. Melanie defined the Blockchain as an transparent and decentralized distributed database for value exchange\cite{swan2015blockchain}. Mougayar\cite{mougayar2016business} believed that Blockchain Technology was value transaction between individuals from a business perspective, a decentralized and de-trusted distributed database from a technical perspective, and a legal perspective that does not require transaction verification by a third-party intermediary. The People's Bank of China defines Blockchain as A technology that is jointly maintained by multiple parties, uses cryptography to secure transmission and access, and enables data consistency, tampering and repudiation prevention, in "Financial application of blockchain technology-Evaluation rules" \footnote{https://www.cfstc.org/bzgk/gk/view/yulan.jsp?i\_id=1866\&s\_file\_id=1787. Last accessed: 2022-04-15.}. There will be diverse understandings from different perspectives of technology. From a network perspective, the Blockchain provides a stable and efficient peer-to-peer network model for distributed data storage. From a structural perspective, the Blockchain is a sequential linked data structure. From the perspective of data storage, Blockchain is a distributed database that both data storage and generation are distributed. From a cryptographic perspective, the Blockchain uses an ellipse Curve digital signature algorithm.

\subsection{Blockchain Structure and Core Technology of Blockchain.} 
    \begin{figure}[h]
        \centering
        \vspace{-0.5cm}
        \setlength{\abovecaptionskip}{-0.1cm}
        \setlength{\belowcaptionskip}{-0.1cm}
        \includegraphics[width=\linewidth]{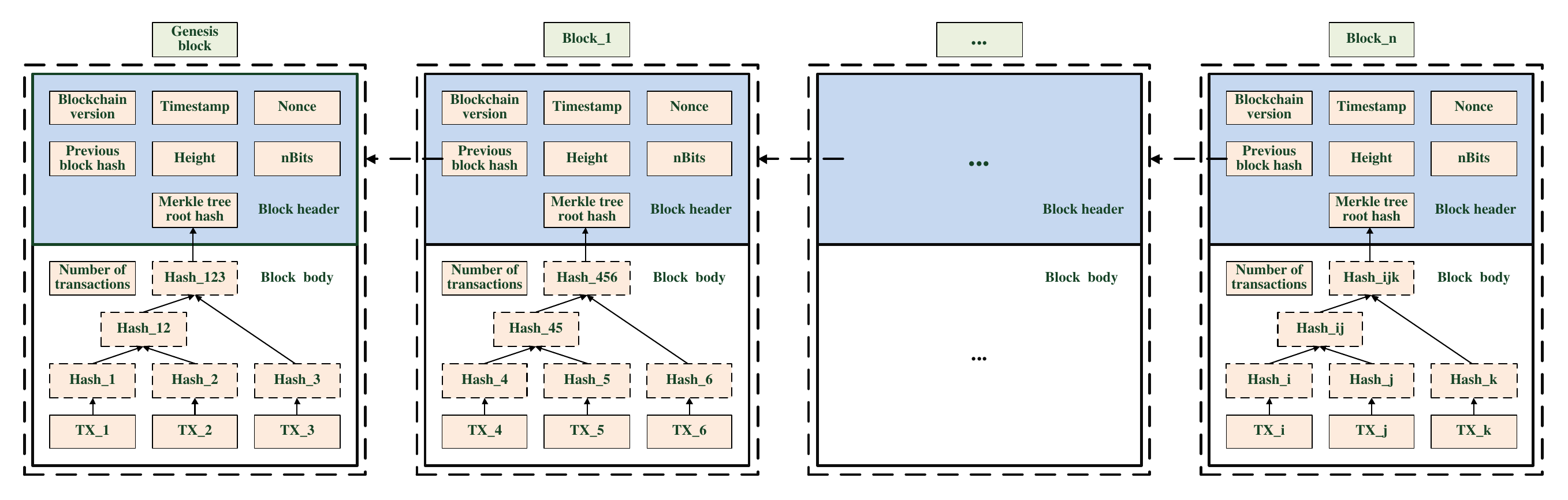}
        \caption{Blockchain structure. Each block points the immediately previous block via previous block hash in the current block header, except the first block which is called genesis block in the Blockchain}
        \label{fig:Blockchain-structure}
    \end{figure}

    \begin{figure}[h]
        \centering
        \vspace{-0.4cm}
        \setlength{\abovecaptionskip}{-0.0cm}
        \setlength{\belowcaptionskip}{-0.3cm}
        \includegraphics[width=0.8\textwidth]{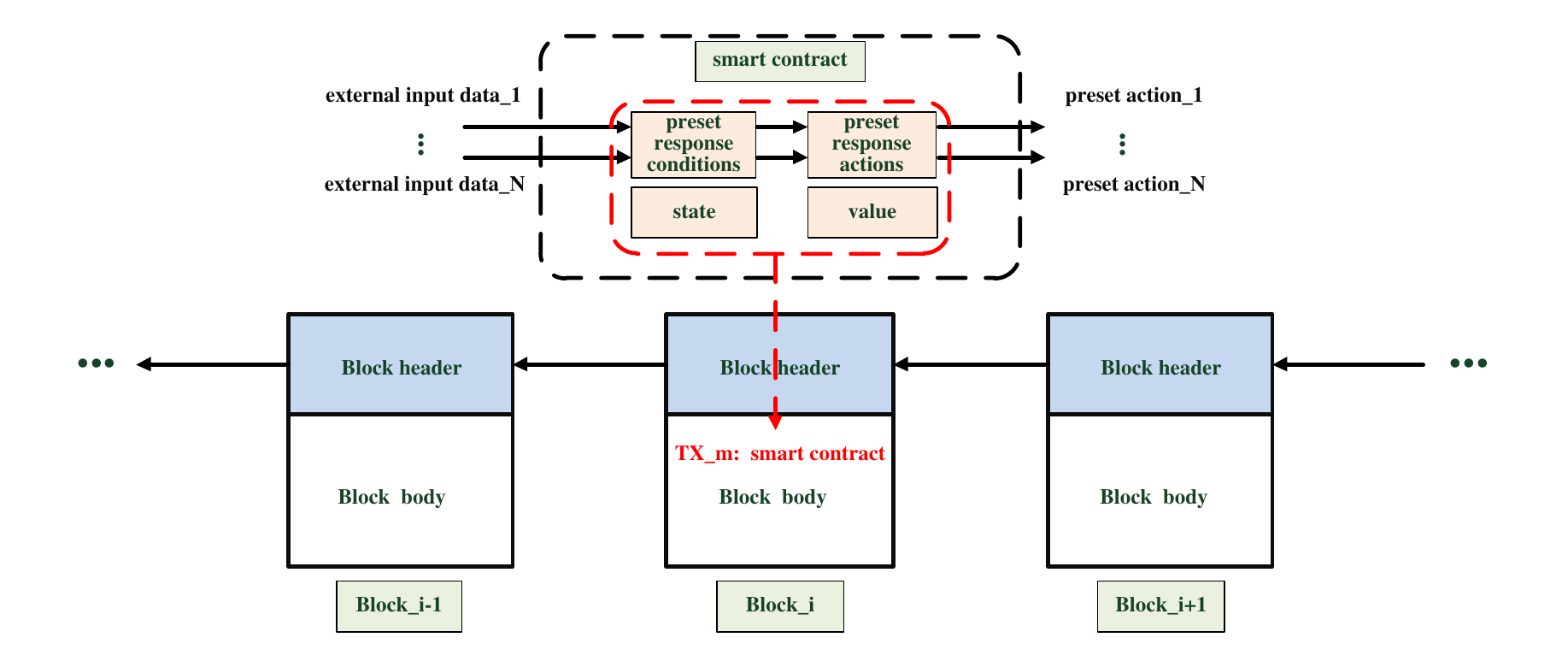}
        \caption{Smart contract model. Smart contract is event-driven, stateful and thresholded, shared computer programs deployed on the Blockchain that enable data processing and automatic execution of triggered actions set by contract rules[53]. First a node completes the deployment of the smart contract and broadcasts it to the consensus nodes for deployment and verification, the result of the successful verification is written to the block as a record. Then, the smart contract can perform its function}
        \label{fig:Smart-contract}
    \end{figure}
The Blockchain consists of a sequence of blocks, each block contains a part of transaction records, and each block chain includes a complete transaction records. A block is composed of block header and block body. 
The block header generally entails: Block version, Previous block hash, Merkle tree root hash, Height, Timestamp, Nonce, nBits. The block body entails transactions and number of transaction.The maximum transaction number indicates how many transactions the block can contain, which depends on the size of the block and the size of each transaction\cite{zheng2018blockchain}. The core technology of Blockchain is not an innovative technology, but a combination of several technologies\cite{YuxiangHuang2018Research}. The core technologies of Blockchain mainly include the following technologies: Distributed storage technology, Encryption technology, Peer-to-peer network technology, Incentive mechanism technology, Smart Contract Technology. Figure. { \ref{fig:Blockchain-structure}} illustrates an example of a Blockchain and Figure. {\ref{fig:Smart-contract}} shows a smart contract model.

\subsection{Research Directions of Blockchain Essential Technology.}
The research direction of blockchain is divided into two major categories by us: the research on the essential technology of blockchain and the research on the combination of blockchain with other computer technologies. 

\subsubsection{Storage}
\ \newline
If stated in a short sentence, blockchain is a distributed storage shared ledger. The study on blockchain storage plays an important role in blockchain research field. The highly redundant storage of blockchain data, which each node stores a complete copy of the data, certainly ensures the openness and transparency of data, but it causes huge storage pressure. At the same time, the huge amount of real-world transaction data also challenges the storage capacity of blockchain for practical applications. Many studies on blockchain storage are conducted. In terms of off-chain storage, Saqib Ali et al.\cite{ali2018blockchain} designed a blockchain-based data storage and access framework using blockchain and distributed hash table(DHT), the metadata of files was stored on the blockchain, while the actual files were stored off-chain at multiple locations through the DHT using PingER (worldwide end-to-end Internet performance measurement project) monitoring agent's peer-to-peer network. In terms of on-chain storage, Huan Chen et al.\cite{chen2019sschain} introduced SSChain, which used an incentive mechanism to ensure that there was no need to periodically reorganize the shard network, avoiding data migration. To reduce the storage burden of the shard, SSChain generated a special block each cycle as a checkpoint, and the state of the blockchain was stored in a sorted Merkle tree for this cycle, so that all transactions before the checkpoint were no longer stored. Currently, blockchain storage technology focuses on challenges such as how to reduce blockchain storage consumption or how to expand the off-chain storage space.

\subsubsection{Peer-to-peer}
\ \newline
Peer-to-peer network lays the foundation of blockchain system. One of the basic starting points of blockchain is decentralization, and peer-to-peer network has the property of state equality among nodes which is exactly in line with the demand of decentralization. Yifan Mao et al.\cite{mao2020perigee} proposed an attractive blockchain peer-to-peer protocol, called Perigee, which had the advantage of computationally lightweight, adversary-resistant, and compatible with the selfish interests of peers, in response to the message propagation latency problem inherent in the underlying peer-to-peer network of the blockchain. Perigee was a decentralized protocol that adaptively determined which neighbors a node should connect to based on its past interactions with its neighbors. Nodes Trade-off between keeping well-connected old neighbors and exploring new neighbors with potentially better connectivity, a node quantified its interactions with neighbors by looking at block arrival times. Selecting neighbors by block arrival time automatically adjusted to the heterogeneity of link latency, block verification latency, and node bandwidth. Zhaoyang Yu et al.\cite{yu2018survey} believed that in the peer-to-peer network of blockchain, nodes had anonymity, which brought trust and selfishness problems, so they proposed to improve the security and sharing of peer-to-peer network from consensus and incentive model. There are much more work studying blockchain peer-to-peer networks\cite{omar2021scalable, tenorio2018open, wu2005optimal}, and we just typically introduce work related to propagation latency , as well as anonymity trust and incentives.

\subsubsection{Encryption}
\ \newline
In the decentralized, transparent and information-sharing blockchain architecture, how to ensure data security and privacy is a matter of great concern, such as on-chain data privacy, data origin authentication, so encryption is also an important research direction of blockchain. The Encryption direction is some cryptographic algorithms that facilitate the use of blockchain. In terms of homomorphic encryption algorithms, Li Chen et al.\cite{chen2019privacy} proposed a Paillier homomorphic encryption algorithm on the Fabric blockchain platform to achieve ciphertext storage of private information and improve protection against external network attacks. It used homomorphic additivity of Paillier algorithm to statistic ciphertext data for improving information privacy protection capabilities. In terms of zero knowledge proof algorithms, Zihang Cao et al.\cite{cao2021design} proposed a decentralized key distribution system that requested and distributed content keys without mutual trust, and used zero-knowledge proof technology to prove the correctness of encrypted content keys to others without revealing it. There are other encryption algorithms, such as attribute-based encryption. Xiaodong Zhang et al.\cite{zhang2021data} proposed a data sharing architecture based on a blockchain system and Ciphertext-Policy AttributeBased Encryption(CP-ABE), called ThemisABE, which had one-to-many data encrypting privacy protection and fine-grained user control. In ThemisABE, the key generation algorithm with joint computation of multiple nodes was improved, any party could be an authorized party, and the CP-ABE algorithm generated keys for data users with corresponding attributes through an arbitrarily chosen set of nodes in the blockchain system. The off-chain computation was used to improve the scalability and privacy of the system, and different types of computation logs were recorded on the chain.

\subsubsection{Consensus mechanism}
\ \newline
Consensus mechanism is considered as the soul of blockchain technology, so a lot of research works on consensus mechanism has emerged. The current blockchain consensus is divided into two main categories\cite{YizhongLiu2019consensus}: permission consensus and permissionless consensus. One of the major categories is permission consensus mechanisms. permission consensus mechanism means that in a blockchain network, nodes need to be permitted to join the network, and then a distributed consistency algorithm is run among nodes to achieve consensus on data, and complete the consensus process to generate and update the blockchain in the network. Typical blockchain permission consensus mechanism researches are Hyperledger\cite{androulaki2018hyperledger}, DFINITY\cite{hanke2018dfinity}, PaLa\cite{chan2018pala}, etc. Their core algorithm is mainly Byzantine Fault Tolerance(BFT). Another major category is permissionless consensus mechanisms. In a permissionless network, all nodes can join and leave the network at any time without permission, the number of nodes in the network changes dynamically, and the process of block proposer election, block generation and node verification to update the block chain is done by a specific consensus algorithm. One of the blockchain consensus mechanisms based on Proof-of-Work(PoW) is to compete for block bookkeeping rights by nodes completing a certain amount of computation (e.g., finding the solution to a specific puzzle). This competition for bookkeeping rights is recognized by all nodes in the network, and non-bookkeeping nodes need to complete the verification of new blocks and complete a new round of competition for bookkeeping rights on this basis. The typical research works based on PoW consensus mechanism are Bitcoin\cite{nakamoto2008bitcoin}, Ethereum\cite{buterin2013ethereum}, Bitcoin-NG\cite{eyal2016bitcoin}, GHOST\cite{sompolinsky2015secure} and IOTA\cite{popov2018tangle}, etc. Another permissionless one is the consensus mechanism based on Proof-of-Stake(PoS). Those who have stakes are responsible for bookkeeping instead of PoW. The probability of becoming the producer of the next block is determined according to the proportion of stakes owned by the user. Nodes are randomly selected as the blocker. The higher the proportion of stakes owned, the greater the probability of becoming the blocker. The typical research works based on PoS consensus mechanism are PPCoin\cite{king2012ppcoin}, Snow White\cite{daian2019snow}, Ouroboros\cite{kiayias2017ouroboros}, etc. Another type of permissionless consensus mechanism is the committee consensus mechanism, which use PoW or PoS to elect specific committee members and committee leaders, and run a distributed consensus mechanism within the committee, and reconfigure the committee after the block is generated. For example, the representative works had PeerCensus\cite{decker2016bitcoin}, Omniledger\cite{kokoris2018omniledger}, RapidChain\cite{zamani2018rapidchain}, etc.
\subsubsection{Incentive mechanism}
\ \newline
The blockchain incentive mechanism can keep nodes honest and motivated to participate in transaction consensus and validation. The existing interesting works on incentive mechanism can be roughly divided into three categories: computation, storage and transmission. The most typical example of computation is the Bitcoin system using the PoW mechanism, which miner nodes were rewarded with the appropriate Bitcoins by calculating the appropriate hash value\cite{nakamoto2008bitcoin}. A better known storage incentive mechanism is Filecoin\cite{ProtocolLabs2012Filecoin}\footnote{https://filecoin.io/filecoin.pdf}, Filecoin was a blockchain protocol token that acted as an incentive layer on InterPlanetary File System(IPFS), Filecoin used Proof-of-Space-time(PoSTs) and PoRep(Proof-of-Replication) to incentivize data storage and retrieval operations. Clients payed to data requirements, while storage miners earned tokens by offering storage and retrieval miners earned tokens by serving data. In the transmission, Xu Wang et al.\cite{wang2021incentivizing} found that the existing blockchain incentive mechanism was not very beneficial to the ordinary nodes responsible for the main message propagation, so it proposed a relay incentive scheme based on the UTXO model to expedite message propagation. A relay node selected the next relay node from its neighbor nodes to complete the propagation of transaction messages, from the sending node to the relay node and then to the receiving node, all the processes were based on the UTXO model to generate certain propagation rewards. In this way, to promote the ordinary nodes to get the due benefit when relaying the propagation information. Incorporating economic incentives into the blockchain technology system promoted a virtuous cycle and development of the blockchain ecosystem\cite{huang2019survey}.

\subsubsection{Smart contract}
\ \newline
A smart contract is a set of computer instructions that are automatically executed according to rules. The combination of blockchain and smart contract is considered as “the next big thing”\cite{abdelhamid2019blockchain}. We discuss this research direction from the four main categories below: security promotion, privacy protection, computation reduction, efficiency improvement thanks to the contribution of research conducted worldwide\cite{hewa2021survey}. Security is the first vital consideration for smart contracts as it directly affects the entire blockchain ledger. More than the other three categories, this category of research is the most\cite{nikolic2018finding, liu2018reguard, jiang2018contractfuzzer, grech2018madmax, tsankov2018securify, grishchenko2018semantic, kalra2018zeus, delmolino2016step}. Kevin Delmolino et al.\cite{delmolino2016step} documented some important experiences about smart contract programming design, and the authors presented common security errors and the ways to fix/avoid them, while advocated best practices. Privacy and sharing seem to be contradictory, but sharing transaction data between peer nodes in the blockchain does not mean information leakage. In fact, what is shared is encrypted data with in-built transparency, or only part nodes are visible. The relationship between privacy and sharing is like the contradiction theory of dialectical materialism. One of the representative works on privacy protection is for example\cite{kosba2016hawk}, Ahmed Kosba et al. presented Hawk, a decentralized smart contract which allowed programmers to write private smart contracts intuitively without implementing cryptography, and the compiler automatically generated an efficient cryptographic protocol using cryptographic primitives, reaching transactional privacy from the public's perspective. They were pioneers to formalizing the blockchain model of cryptography. The computational overhead will be very high if smart contracts are not optimized enough, it will impose additional costs for blockchain users. According this insights, Ting Chen et al.\cite{chen2017under} analyzed that under-optimized Ethereum smart contracts cost more gas and the Solidity compiler failed to optimize the gas-costly programming model.So the authors developed GASPER, a innovative tool that automatically found gas cost patterns in7 gas-costly patterns by analyzing the bytecodes of smart contracts. This would facilitate the optimization of smart contracts. Efficiency improvement is one of the themes pursued by blockchain technology. Parwat Singh Anjana et al.\cite{anjana2019efficient} added concurrency to the execution of smart contracts, to achieve better efficiency and higher throughput. This work developed a framework for concurrently executing smart contract transactions efficiently using optimistic Software Transactional Memory systems(STMs). The STMs executed the smart contract transactions concurrently using Basic Time stamp Ordering(BTO) and Multi-Version Time stamp Ordering(MVTO) protocols. To avoid getting a different final state, transactions executed concurrently in different threads generated block graph for validation. The experimental results show the exciting performance of this concurrent model.

\subsubsection{Cross chain}
\ \newline
In the face of the proliferation of blockchain projects, cross-chain technology has been entrusted to achieve interoperability between blockchains, ultimately solving the problem of information island and forming global information interconnection. Referring to\cite{li2019research}, we summarize four types of cross-chain technologies according to cross-chain mechanisms: Hash-locking, Sidechains/relays, Notary schemes and Communication protocol clusters. Hash-locking is a set of operations with the same trigger set on chain A and chain B, generally the revelation of the preimage of a particular hash\cite{buterin2016chain}. Among the many cross-chain technologies based on the Hash-locking, The work of TierNolan et al.\footnote{https://bitcointalk.org/index.php?topic=193281.msg5116279} is one of the most representative ones, as it is the earliest prototype of blockchain Hash-locking. The author TierNolan et al. proposed atomic transfers conception to let both sides of the transaction set up a contract script on the Bitcoin blockchain, using whether to know the preimage of a certain hash value as the contract trigger condition, the preimage was randomly generated before the transaction, and combined with a series of contract locking and unlocking process designed by the scheme, to achieve cross-chain transactions occurring at the same time or not occurring at the same time. The atomic transfers was later improved and called Hash-locking which one of major technical means of cross-chaining. The Sidechains/relays is a blockchain system that can validate or access another blockchain data. Suppose a smart contract on chain B wants to know the value of a particular state on chain A. The smart contract on chain B accesses a block header of chain A and verifies it. Once the relay has verified the block header, the relay can then verify any desired state value by verifying a single branch of the Merkle tree against the block header\cite{buterin2016chain}. Adam Back et al.\cite{back2014enabling} first introduced the concept of side chains in 2014, which could enable cross-chain transfer of assets between parent chain and side chains without affecting the parent chain. This system allowed for easier interoperability and avoided the liquidity shortage and market volatility associated with new currencies. The Notary schemes refers to a group of parties agree to act on chain B when certain events occur on chain A. Interledger was originally proposed by Stefan Thomas et al.\cite{thomas2015protocol} in 2015 as the cross-chain representative of the notary mechanism. The early version of Interledger used the notary mechanism to realize the connection of different types of ledgers. The Interledger protocol was used for the interaction of different cross-chain systems. Cross-chain transactions were transmitted through the connector. Before the final transaction was completed, each blockchain on the transmission line would escrow and lock the funds of the sender in each link. The participants would select a group of notaries to coordinate the transaction. Communication protocol clusters refer to a cross chain technology that realizes blockchain access mainly by specifying a series of communication data formats and protocol specifications. We like to use the Aion project\cite{spoke2017aion} as an introduction to illustrate cross-chain technology for communication protocol clusters, Matthew Spoke et al. defined a multi-layer blockchain network architecture that allowed different blockchain systems to access though communication protocols in 2017. It specified the originating network, destination network, routing information, Merkle proof and so on. In the inter-chain transaction method, Blockchains were connected and communicated through bridges, and the bridges had their own verification network. They designed the Aion Virtual Machine(AVM) to abstract the blockchain logic and provide an environment for running applications.

\subsection{Applications on the combination of blockchain with other computer technologies.}
Another major category is the integration of blockchain with other computer technologies to jointly serve the Supply Chain Finance field. We briefly list some interesting related applications and summary from the following main aspects.

\subsubsection{Blockchain+ Data analysis}
\ \newline
The Blockchain is essentially a distributed database with its own unique data characteristics: data anonymity, sharing and privacy, block data relevance in chronological order, non-tampering and so on. The mining and analysis of Blockchain data is not only the inherent requirement of the characteristics of its data attributes, but also the actual needs of the Blockchain should be implemented in the actual industry. As the literature\cite{huang2021survey} said, the work of Blockchain data analysis is still relatively little. Here we introduce a relatively highly relevant review\cite{weili2018blockchain}, the authors divided the Blockchain data into transaction data and contract data, and the contract data included code data and transaction data generated by the trigger. Then, they summarized the current research status and future challenges of Blockchain data analysis from seven aspects: entity identify, privacy disclosure risk analysis, network portrait, network visualization, transaction pattern recognition, market effect analysis, and illegal behavior detection.

\subsubsection{Blockchain + Internet of Things}
\ \newline
The advantages of data integrity, security, privacy, security, privacy, sharing, decentralization and other advantages brought by Blockchain technology have attracted the attention of many Internet of Things researchers\cite{ali2018applications, panarello2018blockchain, dai2019blockchain, wang2019survey, lao2020survey, peng2021security}. One of the jobs we admire is\cite{lao2020survey}, Laphou Lao et al. conducted a systematic investigation into the key components of the IoT Blockchain. First, an overview of the architecture of the popular IoT Blockchain system was given by analyzing the network structure and protocol. Then it compared the consensus algorithms and communication protocols of various IoT Blockchains, demonstrated its advantages in the application of IoT Blockchains. On this basis, a new universal IoT Blockchain framework was proposed. At the same time, the authors presented a traffic model of the Internet of Things-Blockchain system on the basis of analyzing the traffic model of P2P and Blockchain systems.

\subsubsection{Blockchain + Artificial Intelligence(AI)}
\ \newline
As with Blockchain technology, artificial intelligence is also very much hyped as an innovative technology. In general terms, an important function of AI technology is to allow computers to learn, make decisions and control matters like humans do. This re-emergence of AI is supported by massive data resources. We argue that Blockchain has a reliable data storage function, while AI technology has a powerful data processing and decision making capability. The combination of Blockchain and AI technologies can help alleviate each other's technical challenges. On the one hand, AI technology can leverage Blockchain to obtain credible training data, which can result in more accurate models. On the other hand, Blockchain can utilize AI technology to improve data processing capabilities and extend the functionality of smart contracts. A couple of review literatures discussed the combination of Blockchain and AI, and the implications of such integration\cite{salah2019blockchain, zheng2019blockchain, liu2020blockchain}. One of the classical review is\cite{salah2019blockchain}, Khaled Salah et al. presented a detailed survey on the role that Blockchain plays in the context of AI. Firstly, the authors given overview of Blockchain and how Blockchain features can assist in AI such as enhanced data security, improved trust on robotic decisions, collective decision making, decentralized intelligence, high efficiency. In addition, they highlighted the primary advantages of this integration. Then, the article presented a detailed taxonomy of Blockchain types, decentralized infrastructure, consensus protocols, along with existing decentralized AI operations and applications. They discussed works reported in the article on how Blockchain-enabled AI enhance the various performances of data and algorithms in AI applications. This review provides an important basic reference for subsequent research as the summarized outcomes of related work and the challenges raised.

\section{Applying Blockchain to Supply Chain Finance} \label{title:Blockchain-SCF}
\subsection{literature research content and trend analysis.}

In order to get an approximate understanding about the application for Blockchain technology in the field of Supply Chain Finance, we mainly search relevant papers from two comprehensive databases: Web of Science and Scopus. The purpose of selecting these two popular literature data is to find the main current hot-topics of Blockchain-based Supply Chain Finance in such a way as to inspire us to propose forward-looking future research directions, rather than pursuing a comprehensive review. We hvae searched 98 papers in Web of Science core collection database using ( Blockchain (Topic) AND Supply Chain Finance (Topic))\footnote{https://www.webofscience.com. Last accessed: 2022-03-21} , and 230 papers in Scopus using (TITLE-ABS-KEY(Blockchain) AND TITLE-ABS-KEY(supply  AND chain AND finance))\footnote{https://www.scopus.com. Last accessed: 2022-03-25}. 248 papers are obtained after removing duplicate papers for 98-Web of Science and 230-Scopus papers, and we evaluate whether each paper is related to Blockchain (BC) or Supply Chain Finance (SCF). 

As can be seen from the Table {\ref{tab:scf_bc_retrieved}}, the first major category of the literatures we investigated has three literatures that are neither related to SCF nor blockchain\cite{salvo2021security, li2017progress, chae2022technologies}, which are associated with Industrial Internet, Fintech and Industry 4.0, respectively. The second major category has 8 papers, which are related to Supply Chain Finance but not to blockchain. These works are more or less related to Supply Chain Finance and are broadly subdivided into supply chain operations\cite{luo2022operational, bal2021supply}, financing models\cite{tang2021manufacturers, choi2020financing, hu2021value, wandhofer2019financing}, surveys\cite{tseng2021comparing} and editorials\cite{choi2020editorial}. The third major category is not related to Supply Chain Finance but to blockchain, and this category has the largest number of 159 articles. Interestingly, we found that 47 of them are surveys, involving the essence of blockchain\cite{tolmach2021survey}, intelligent transportation\cite{jabbar2022blockchain}, smart cities\cite{ridjic2022implementation}, finance\cite{franco2014understanding}, supply chain\cite{gurpinar2021current}, healthcare\cite{vervoort2021blockchain}, etc. It shows that people are very much looking forward to the present and future development of blockchain, trying to apply Blockchain technology is applied to the next stage of planning and development in various industries, which is reflecting the phenomenon of a lot of research and summarization work in the initial stage of blockchain application. It shows that people are very much looking forward to the present and future development of blockchain, trying to apply Blockchain technology is applied to the next stage of planning and development in various industries, which is reflecting the phenomenon of a lot of research and summarization work in the initial stage of blockchain application. As for the remaining 112 papers in the third major category, the main hot directions of the literature are not only related to the nature of blockchain introduced in the previous section\cite{dai2020research} and combining with other technologies\cite{liang2020pbft}, but also reflected in several application industries, such as education\cite{zhu2021research}, construction\cite{nanayakkara2021blockchain}, COVID-19\cite{singh2023boss}, carpooling system\cite{vazquez2021towards}, and so on. Because there are too many fields and papers involved, and the focus of our research is not the direction of blockchain application, we will not list these references one by one here. Finally, these remaining 78 are literature related to both blockchain and Supply Chain Finance, including three surveys: survey\cite{trautmann2021blockchain} mainly discussed the problem of inefficiency of traditional Supply Chain Finance, and briefly analyzed how blockchain could solve the problems of opaque information, asymmetry and inefficient process. The survey\cite{huang2021application} briefly introduced that blockchain could realize the function of digitizing assets, reducing risks and alleviating information asymmetry, and broadly suggested promoting the practical application of blockchain, strengthening cooperation and new technology regulation. And the survey\cite{li2019simple} mainly focused on the summary of risk management in Supply Chain Finance. These reviews summarized and suggested more at the macro level. Compared with them, our survey focuses more on the refinement of technology-industry combination, discusses the overall status of blockchain application to supply chain in more detail, and discusses its research direction from the perspective of computer science. 
    \begin{table*}[!t]
        \centering 
        \caption{Classification table of the retrieved articles}
        \label{tab:scf_bc_retrieved}
        \renewcommand{\arraystretch}{3}
        \begin{tabular}
        {m{1.2cm}<{\centering}
        |m{1.2cm}<{\centering}
        |m{1.2cm}<{\centering}
        |m{1.6cm}<{\centering}
        |m{8cm}<{\centering}}
         \toprule
            \textbf{Related to SCF} & \textbf{Related to BC} & \textbf{Number of articles} & \textbf{Years} & \textbf{Main aspects} \\
            \hline
            \ding{53} & \ding{53} & 3 & 2017, 2021 &  Industrial Internet\cite{salvo2021security}, Fintech\cite{li2017progress}, and Industry 4.0\cite{chae2022technologies} \\
            \hline
            \ding{51} & \ding{53} & 8 & 2019\textasciitilde 2022 & Supply chain operations\cite{luo2022operational, bal2021supply}, financing models\cite{tang2021manufacturers, choi2020financing, hu2021value, wandhofer2019financing}, reviews\cite{tseng2021comparing} and editorials\cite{choi2020editorial}\\
            \hline
            \ding{53} & \ding{51} & 159 & 2014\textasciitilde 2022 & Survey: blockchain essence\cite{tolmach2021survey}, smart transportation\cite{jabbar2022blockchain}, smart city\cite{ridjic2022implementation}, finance\cite{franco2014understanding}, supply chain\cite{gurpinar2021current}, healthcare\cite{vervoort2021blockchain}, etc.; Research directions: blockchain essence\cite{dai2020research}, combination with other technologies\cite{liang2020pbft}, education\cite{zhu2021research}, architecture\cite{nanayakkara2021blockchain}, COVID-19\cite{singh2023boss}, carpooling system\cite{vazquez2021towards}, etc.\\
            \hline
            \ding{51} & \ding{51} & 78 & 2018\textasciitilde 2022 & Survey\cite{trautmann2021blockchain, huang2021application, li2019simple}, Risk Management\cite{gao2018real, fu2019big, wang2021research, salman2019reputation, liu2021hybrid, salman2018probabilistic, xie2021risk, XiaoyuDuan2020mastersthesis}, As a Service Platform\cite{wang2021research-Platform, li2021optimization, zhao2016coordination}, Financing Model Innovation\cite{du2020supply, yuyan2020role, wang2021research-Platform, li2021optimization, jiang2020application, yang2021blockchain},Credit Management\cite{ma2022supply, chen2021applying, xu2021research, yu2021financing}, Data Management\cite{jiang2021privacy, zhang2021research, liu2021mitigating, li2020fabric}, Smart Contract Application\cite{zhang2021internet, chen2020blockchain, aimin2019intelligent, zhang2021supply, yan2020confidentiality}, etc.\\
            \hline
        \end{tabular}
    \end{table*}

In addition, according to the analysis of the application bright-spots of 78 well-relevant articles on the combination of blockchain and Supply Chain Finance, we divide several main application hot-spots and the percentage of related articles: Risk Management accounts for 15\%, Service Platform accounts for 11\%, Financing Model accounts for 16\%, Credit Management accounted for 6\%, Smart Contract Application accounted for 6\%, Data Management accounted for 8\%, and Others accounted for 38\%. Based on this statistical analysis, we have identified 6 application hots-pots, and will introduce and evaluate these current hot-topics in the subsequent sections.

\subsection{Application hot-topics.}
According to 78 related papers we have analyzed 6 categories of application hot-topics: Risk Management, Service Platform, Financing Model, Credit Management, Smart Contract Application, Data Management. They are briefly described below.

\subsubsection{Risk Management}
\ \newline
The applications of Blockchain and Supply Chain Finance in risk management fall into three main categories. 

The first category is to regard Blockchain as a trusted data storage tool\cite{gao2018real, fu2019big, wang2021research}. The data generated by the supply chain, such as capital flow, logistic flow, information flow and business flow, are uploaded onto the Blockchain, and the effects of data anti-counterfeiting, data transparency and data consensus validation can be realized in the Blockchain. The Blockchain alleviates the difficult problem of asymmetric supply chain data information, and facilitates financial institutions (such as banks and credit companies) to conduct comprehensive and timely risk assessment of financing companies (such as suppliers and retailers) in the supply chain, it also promotes the improvement and innovation of financing models. Such credible data reduces the difficulty when assessing risks, saves assessment costs, and makes the assessment results more credible. For example, Gao B et al.\cite{gao2018real} argued that Blockchain was a distributed database and the information generated by each node in Supply Chain Finance would be broadcast to the whole Blockchain network for information sharing, saving information request and transmission, improving the efficiency of Supply Chain Finance operations, and reducing the risks caused by information asymmetry, information infection, and human manipulation of data. The authors avoid bank losses through timely capital replenishment by comparing real-time data provided via Blockchain with pledge loss thresholds, and further discussed that Blockchain could help secondary and tertiary suppliers and retailers to participate in Supply Chain Finance and obtain loans, especially Small and Medium Enterprises. Fu Y et al.\cite{fu2019big} researched the storage model of data on Blockchain, they composed subsection-blocks of similar supply chain data, different subsection-blocks were logically connected, each subsection-block was shared by an authentication center, and similar nodes stored the same class of subsection-blocks. Meanwhile, consensus mechanism was employed in the supply chain to reach reasonable demand and cooperation information jointly through multiple rounds of voting. 

\begin{figure}[h]
        \centering
        \vspace{-0.3cm}
        \setlength{\abovecaptionskip}{-0.1cm}
        \setlength{\belowcaptionskip}{-0.3cm}
        \includegraphics[width=\textwidth]{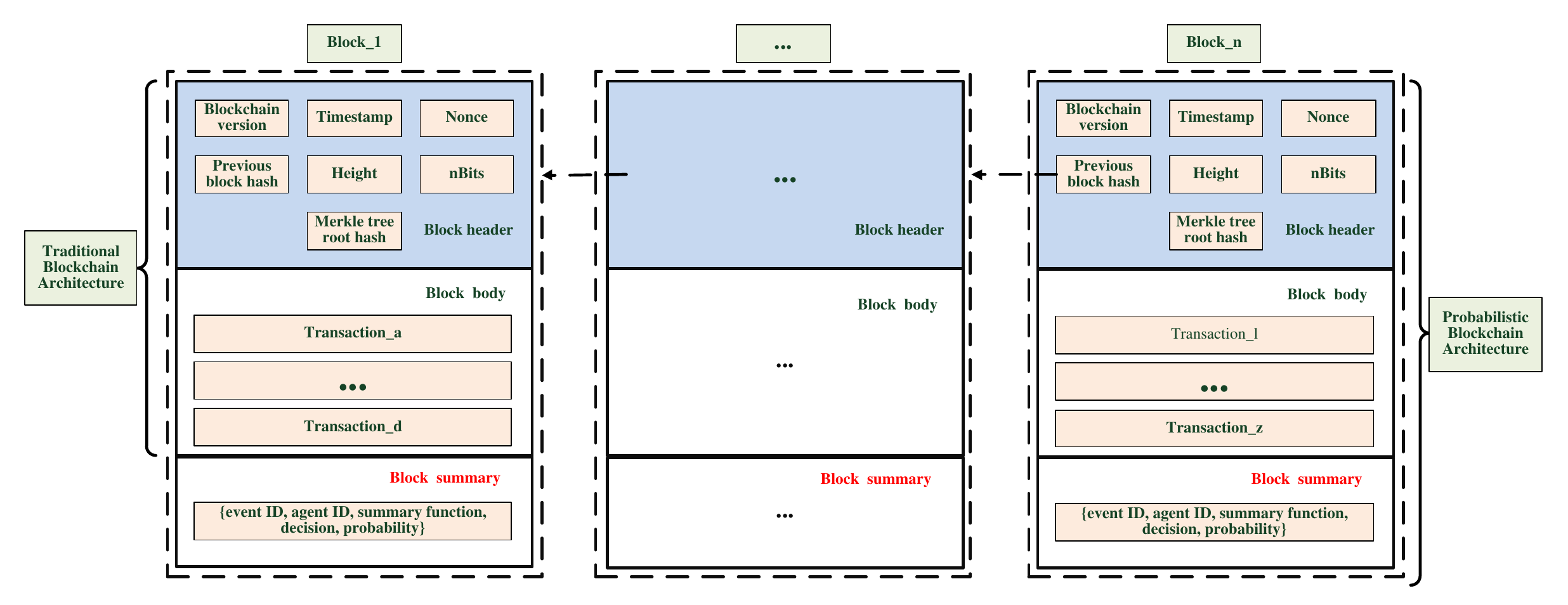}
        \caption{The probabilistic blockchain architecture and traditional blockchain architecture}
        \label{fig:probabilisticBlockchain}
    \end{figure}

The second major category is the transformation of Blockchain from a data storage tool into into a Blockchain computing that both stores and collaboratively processes data\cite{salman2019reputation, liu2021hybrid, salman2018probabilistic}. Such researchers were not satisfied with Blockchain being used only as a means of data storage, but explored the possibility of transforming data into knowledge, expanding Blockchain storage to Blockchain computing, and moving from on-chain information storage to decision making in the era of big data and machine learning. The expansion and extension of Blockchain could enable it to be better combined with other technologies, such as data mining and artificial intelligence, providing greater possibilities and influence on the implementation of technology applications. The work of Salman T et al.\cite{salman2019reputation} was admired by us, they intended to extend the use of Blockchain for computing, so they introduced a type of Blockchain, called probabilistic Blockchain, which could provide a distributed and immutable database allowing securely to store individual decisions and allowing consensus decision. Compare to traditional Blockchain, which contained a set of distributed nodes maintaining the database and were connected via a network, the probabilistic Blockchain allow transactions to be simply the views of users, not just deterministic events or transactions only. In addition to transaction records, probabilistic Blockchain included a block summary that consisted of events, agents(nodes), decision function(summary function), decision and response event probability. Here Blockchain computing was used for block summary. The probabilistic Blockchain structure is shown in Figure.{\ref{fig:probabilisticBlockchain}}. Then the authors extended the knowledge-based Blockchain on the basis of the probabilistic Blockchain paradigm, knowledge-base Blockchain was potentially valuable in building efficient risk assessment respect, and designed a reputation system suitable for such Blockchain with which to perform malicious node detection. Here is another work on Blockchain computing that deserves to be presented\cite{liu2021hybrid}, Liu J et al. argued that although Blockchain addressed the problems of incorrect information between enterprises, lack of visibility of the transaction process, and possible joint fraud in the core enterprise model, there were still problems such as inconvenient information verification, data falsification, and difficulty in balancing efficiency, security, and cost, so they proposed a hybrid chain model that combines PANDA (public chain-based consensus algorithm) public chain-based consensus algorithm) and X-Alliance (coalition chain-based consensus algorithm) hybrid chain model, which could process each account's transactions in parallel and asynchronously with other unrelated accounts in the network, making the enterprise's transaction authenticity review, risk assessment and credit delivery more efficient as well as lower cost and more secure performance.
    
What we define as the third category is to see Blockchain as a risk factor in Supply Chain Finance\cite{xie2021risk, XiaoyuDuan2020mastersthesis}. In this category, Blockchain database or Blockchain computing is not distinguished, but rather as an element of the overall risk control system, considering what impact, threat or benefit it would bring to Supply Chain Finance. One of the representative works we are aware of is a three-level Risk hierarchy model of Supply Chain Finance on Blockchain constructed by Xie W et al.\cite{xie2021risk}, which the secondary indicators consisted of macro-industry analysis, credit risk, supply chain relationship risk, pledge risk, operational risk, Blockchain system risk. It is believed that stability risk, security risk, reliability risk and external macro Blockchain policy factors of Blockchain technology affected the whole Supply Chain Finance risk management system. The scores of each risk influencing factor were obtained by questionnaire and expert consultation, and then the weights of each factor were calculated by combining analytic hierarchy process and fuzzy cognitive maps method to quantify the risk analysis to some extent. Similarly, Duan Xiaoyu et al.\cite{XiaoyuDuan2020mastersthesis} explored the establishment of credit risk evaluation indexes for Supply Chain Finance after the introduction of Blockchain, and added Blockchain technology effectiveness as a factor affecting risk to the previous credit evaluation system, which includes evaluation indexes such as the degree of information sharing, traceability, and cryptographic security.

In our opinion, these current works see blockchain as a new distributed storage tool, and it is simple and easy to use it for Supply Chain Finance risk management. This is because how to ensure the trustworthiness and security of blockchain data is the main focus of professional blockchain technologists and researchers.As an application level, a simple on-chain storage of data is required under the assumption that the blockchain is trusted. Therefore, the current work of Supply Chain Finance based on blockchain is easy to implement. Further consideration of blockchain not only as a storage pool but also as a computational processor is more in-depth work, which is in its infancy and worthy of continued innovative work by experts in blockchain applications, as it can give blockchain more powerful functions to serve the practical needs of the industry. As for considering blockchain as a risk factor, this approach has some justification, because anything new entering the system will inevitably cause changes in response and bring certain uncertainties, thus generating risks. This can provide a reference for the establishment of a more comprehensive risk evaluation system. In addition, the current lack of publicly available validation datasets for blockchain-based supply chain datasets also brings troubles of experimentation and validation to this part of work.
\subsubsection{As a Service Platform}
\ \newline
With distributed, shared and trusted attributes, Blockchain is often regarded as a trustworthy platform for Supply Chain Finance business. Appropriate type of Blockchain(usually federated chain) can be joined to allocate authority, information sharing and business operations among supply chain members. The flow of capital, information, business and logistics in the supply chain can be recorded in the Blockchain, providing evidence for real business operations and transaction verification. Here, Blockchain is like a factory, and the operation of each department in this factory(suppliers, core enterprises, sellers, financial institutions) is recorded in the factory database. This hot-topic application focused on engineering application and management research. Focusing on the service platform work in the engineering category is mainly the service platform architecture design\cite{wang2021research-Platform, li2021optimization, zhao2016coordination}. For instance, a new Blockchain-based supply chain financial service platform was proposed in the literature\cite{wang2021research-Platform}, which was designed to consist of five modules: authority management module, digital asset management module, contracts management module, credit extension management module, and traceability module. The work detailed the operation process of the service platform, showing that Blockchain as the base of the whole new supply chain financial system brought the advantages of smarter and more reliable. Another example is the research work of literature\cite{li2021optimization}, which designed a four-layer architecture from the demand of service platform, and Blockchain as one of the basic layer design played the role of providing logistics information supervision and sharing for the whole platform and accelerating the efficiency of financing. In contrast, the work belonging to the management research category focuses on the impact of Blockchain on Supply Chain Finance, comparing the case with and without Blockchain. This type of work examines the macro-management changes brought about by Blockchain from a management perspective, rather than the specific platform design and implementation. The literature\cite{zhao2016coordination} investigated a single-cycle chain-controlled supply chain based on a Blockchain platform. By comparing a normal platform for Supply Chain Finance with a Blockchain-based Supply Chain Finance platform, the relationship between the utilization rate and default rate of the Blockchain service platform and the loan amount when funds are restricted was found. 

In general, the research work on utilizing blockchain as a service platform for Supply Chain Finance is generally based on the exploration of macro concepts, considering what kind of business improvement can be brought on the basis of blockchain. However, the practice of adopting blockchain as a service platform has not been fully developed, and some application on the ground shows that such work is still in the early stage of experimentation. More research work is needed to follow up on how to enhance the functionality of blockchain platforms and to demonstrate the indispensable ability of blockchain practical applications.
\subsubsection{Financing Model Innovation}
\ \newline
Different from the service platform application, the financing model mainly affects the business link of Supply Chain Finance. As introduced in the previous content, Supply Chain Finance is a tool to help supply chain capital operation, and its links include financing links and supply chain business links, and financing will be guaranteed by assets such as accounts receivable, inventory and prepaid accounts generated in supply chain business. At this point, Blockchain is equivalent to a segment in the assembly line, and together with the supply chain segment and financing segment, it constitutes the new model of Supply Chain Finance. The literature\cite{du2020supply} utilized Blockchain technology innovation to transform the business model of Supply Chain Finance and re-designed the processes and business models of warehouse receipt pledge financing and accounts receivable factoring in Supply Chain Finance, so as to ensure the information symmetry in Supply Chain Finance and realize the controllability of financing business. The literature\cite{yuyan2020role} applied Blockchain technology to the accounts receivable business model of Supply Chain Finance, and analyzed the advantages of applying Blockchain technology to Supply Chain Finance by analyzing the current situation and problems of the traditional accounts receivable business model in order to promote the establishment of the credit system of Supply Chain Finance and facilitate the financing business. 

It is worth emphasizing again that Blockchain in Blockchain-based service platform is used as the foundation of the whole platform to optimize the operation of supply chain financial system, while Blockchain in Blockchain-based financing mode is specific to a certain financing business and used as one of the links to optimize a certain mode of financing business. According to the papers we searched, most of the studies on management aspects in service platforms and financing models are not work in the field of computer applications technology, but many of them propose conceptual models or use management methods\cite{wang2021research-Platform, li2021optimization, yuyan2020role, jiang2020application, yang2021blockchain} to study the changes brought by Blockchain to Supply Chain Finance operations. This lack of research work from a computer technology perspective is challenging and provides a new entry point for future research work.

\subsubsection{Credit Management}
\ \newline
Credit management is one of the major conditions for Supply Chain Finance to carry out financing business, and credit is treated as an important basis regardless of the amount determination before financing, the model selection during financing, or even the risk control of capital after financing. According to the classification of retrieved papers, credit management in Blockchain-based Supply Chain Finance applications mainly includes credit assessment, credit disassembly, credit transmission analysis and credit risk control. Immediately after, a typical work of each of them will be presented here. The literature\cite{ma2022supply} presented a credit evaluation mechanism for enterprises in Supply Chain Finance that combined federal learning with Blockchain technology, referred to as data credit evaluation model. one of the Blockchain modules consisted of supply chain participants and financial institutions to record various processes and results in the supply chain and financing process, and then implemented credit evaluation using the federal learning module and broadcast the results of federal learning on the Blockchain. Based on the results of trust evaluation, financial institutions decided whether to reach financing agreements. And the past credit evaluation results of financing companies and the progress of financing projects could be checked on the Blockchain. The literature\cite{chen2021applying} designed to implement a credit disassembly mechanism with Blockchain technology to reshape the business process of Supply Chain Finance and innovate a Blockchain-based Supply Chain Finance system, which offered a new perspective to address the development of Supply Chain Finance. The literature\cite{xu2021research} analyzed the credit transmission when information was asymmetric and influenced the credit mechanism through core firms' enhanced credit, Blockchain application and automatic execution of smart contracts. The credit perception decay based on information asymmetry could be mitigated when the transaction data of the supply chain was passed to the Blockchain, and the consensus mechanism enabled the participants to record and maintain the data together to achieve a more credible signature check on the transaction chain, which avoided the risk of data fraud and provided credit assurance for the transaction data and ensured the consistency and authenticity of information. As for credit risk, the literature\cite{yu2021financing} explored the effectiveness of enterprise self-guarantees for financing, and this mode of relying on their own credit to obtain financing made financial institutions to bear a certain credit risk, and the analysis of this work suggested prepayment or repayment in advance to reduce the credit risk of corporate customers, especially to allow low credit customers to prepay service fees according to a certain prepayment ratio, and to choose the best financing strategy under different conditions to control credit risk.

According to our statistics, this type of credit management research is not particularly abundant at the moment, but the work on credit assessment, credit disassembly, credit transmission, credit risk, etc. is very significant because it is an important prerequisite for risk management and business cooperation. The current Simulation experiments are also parameter setting and not validation from realistic data sets. Therefore, more credit research methods and validation need to be continued.
\subsubsection{Data Management}
\ \newline
The data management function provided by Blockchain for Supply Chain Finance is inseparable from the attributes of Blockchain itself. Blockchain is a node-consensus, cryptographic, tamper-proof, denial-resistant and decentralized database, which means that Blockchain can provide privacy protection, tamper-proof, data sharing, and data storage functions. Here are also many researches on Blockchain being applied to Supply Chain Finance on data management. In the literature\cite{jiang2021privacy}, a privacy budget management and noise reuse approach based on Hyperledger multi-channel technology and community clustering algorithm in a multi-chain Blockchain environment was proposed to address the query restriction problem of different privacy requirements in Blockchain-based Supply Chain Finance, which enabled querying privacy budget data by establishing historical records and enabled data providers to have a clearer insight into the usage of data. A differential privacy-preserving algorithm based on noise reuse was also applied to generate and reuse noise to ensure noise reuse under the premise of querying the same type and specific data utility. The literature\cite{zhang2021research} presented a cryptographic Blockchain-based method to prevent the key features of supply financial data from being tampered. The data feature evaluation criteria, data key feature mention, and key feature tamper-proof method were adopted to ensure that the features of distributed data were not tampered, and finally the simulation experiments proved the tamper-proof performance of the method. In the literature\cite{liu2021mitigating}, IoT and Blockchain technologies were jointly utilized to mitigate information asymmetry challenges and enhance data sharing. Combined with the analysis of multiple case studies, it was clarified that IoT could mainly improve the information access of traditional banks and Blockchain could facilitate trusted information transformation and sharing. Blockchain itself is a database, and it goes without saying that data storage functions are used in any Blockchain-based application area. However, Blockchain data storage in this part refers to the study of the optimization of storage methods to facilitate the access and query of data in the Blockchain, etc. For example, literature\cite{li2020fabric} proposed a storage model for electronic document data of supply chain financial data, which combined Blockchain and bloom filter storage methods, where the bloom filter structure was introduced as an index for data storage and it was searched first during the query process. Simulation experiments demonstrated that the model proposed in this work had greater improvement than existing models in terms of access efficiency and retrieval efficiency.

As mentioned before, the current blockchain-based supply chain financial data management is still at the stage of simplistic application, and they utilize the essential characteristics of blockchain to realize the security and sharing of supply chain financial data. We believe that blockchain-based supply chain financial data management can be done in a deeper and more fine-grained level in the future. The existing work only explores a certain application challenge, without considering how the existing realistic data can be put into the blockchain and whether it will cause short-term overload data on the chain. Nor has it considered fine-grained data management, such as the consideration of different data users and the management of different types of data. Similarly, in the face of the fact that there is no publicly available blockchain-based Supply Chain Finance dataset, how to build a data platform for free and public access by the majority of scholars is also a work worth doing now and in the future.

\subsubsection{Smart Contract Application}
\ \newline
In the context of Supply Chain Finance, another common function of Blockchain is the automatic execution operation brought by smart contract. A smart contract can be regarded as an intelligent controller on the Blockchain, which consists of trigger conditions, thresholds and trigger actions, and this intelligent controller program is deployed on the Blockchain and recognized by consensus of nodes. The automatic execution of smart contracts replaces manual operation and avoids the risk of some operational mistakes. Especially for the financing business that requires accurate information data, the availability of computers instead of manual operations is exactly what Supply Chain Finance needs.
Smart contract is used for risk management, platform service, financing model innovation, credit management and data management in Supply Chain Finance. The literature\cite{zhang2021internet} focused on Blockchain-based IoT regulatory system and Supply Chain Finance regulatory methods. Smart contracts were utilized to process data, operate asset transactions and manage smart assets so that data traceability could be achieved, thus ensuring data authenticity and security. Here, smart contracts were used in Supply Chain Finance regulation for risk management. In the work of literature\cite{chen2020blockchain}, smart contract got used in Blockchain-driven Supply Chain Finance platform BCautoSCF to automate part of the workflow of Supply Chain Finance so as to reduce human errors and disruption during contract execution. Moreover, the authors intended to work on the full automation of the workflow of Supply Chain Finance supported by smart contracts in the future to further improve the efficiency of transactions. One of the research works in which smart contract has been applied to the innovation of financing models is\cite{aimin2019intelligent}, where smart contract is used in the factoring business area of Supply Chain Finance. This work modeled the automatic execution mechanism of smart contract in three factoring business application scenarios: split transfer of debt certificates, factoring financing of upstream suppliers and payment due from core enterprises, and verified that Blockchain rational nodes would always follow the relevant protocols to execute certain operations automatically. In terms of credit management in Supply Chain Finance, the literature\cite{zhang2021supply} improved the credit mechanism using smart contracts of Blockchain to promote complete credit delivery and alleviate pain points in Supply Chain Finance, such as trust deficit. The characteristics of determinism, consistency, observability, verifiability, Efficient, real time, and low cost were considered in the process of implementing smart contracts. The design of the smart contract subject and the token mechanism in the smart contract facilitated the credit flow and expanded the financing radiation. The last thing to discuss is smart contract and data management. It is well known that Supply Chain Finance has extremely high requirements for data privacy and confidentiality. The literature\cite{yan2020confidentiality} presented a system design of CONFIDE with data transfer protocol and data encryption protocol, and then an efficient virtual machine using Trusted Execution Environment(TEE) to ensure data privacy and confidentiality.The smart contract extension of CONFIDE could be flexible to meet the needs of complex models. CONFIDE, as a plug-in for the Supply Chain Finance platform AntDuoChain\footnote{https://www.ledgerinsights.com/tag/ant-duo-chain/. Last accessed: 2022-03-31.} module, has been proven to be effective in practice.

It can be seen that smart contracts are applied in each of the previous hot directions, which is because its control function is the key interface point between theory and practice. The existing blockchain smart contracts are mechanical and basic applications in the field of Supply Chain Finance. Manually preset trigger conditions and then execute the corresponding manually preset actions. In our investigation, there is no more intelligent smart contract yet, and how to automate the conditions, thresholds and action settings is a challenging and definitely needed task in the future.

\section{Future directions on Blockchain-based Supply Chain Finance} \label{title:future}
Blockchain-based Supply Chain Finance has received more and more attention in recent years for its wide applications. Thus, its future research directions, including the challenges it faces at the moment, highly deserving of discussion. We derive valuable research directions from the challenges.

\subsection{Data Management for Supply Chain Finance.}
\subsubsection{Blockchain-based Supply Chain Finance Dataset Construction}
\ \newline
To the best of our knowledge, there is no publicly available Supply Chain Finance dataset for use by a majority of scholars, let alone a publicly available Blockchain-based Supply Chain Finance dataset. Data is the basis for reliable analysis in Supply Chain Finance. Both transaction verification, risk management and smart contract execution cannot be done without data as a source of information. Based on the computer application perspective, we need a data foundation that can be computed and analyzed to provide a data basis for the model and enhance the credibility for experimental validation.like Elliptic dataset, Bitcoin Tweets dataset for Bitcoin transaction, a certain scale Blockchain-based Supply Chain Finance dataset is also a vital resource for developing advanced, data content-based algorithms for Supply Chain Finance risk control, smart contracts, etc., as well as providing critical training and benchmark data for these algorithms. To construct such a dataset, a feasible method is to to collaborate with financial institutions to co-construct and appropriately anonymize some corporate identity information to protect privacy. In addition, data characteristics, such as type of financing, timeliness, geography, etc., need to be taken into account. In parallel to this, it is necessary to consider the structural form of the data block, for case, a traditional Blockchain or a knowledge-based Blockchain.

\subsubsection{Data privacy protection}
\ \newline
Another research direction in data management is the privacy protection of data at fine grained level. A typical feature of Blockchain as an enhanced version of distributed database is the data encryption feature. Privacy protection itself is also a hot topic of research in Blockchain technology\cite{liu2018research, feng2019survey, wang2020survey, cao2020privacy, zhang2021overview}. In Blockchain networks, the nodes of the whole network need to reach a consensus on the transaction data before the data is uploaded, which means that the data needs to be openly available on the Blockchain network, thus bringing the privacy protection trouble\cite{gao2018real}. Similarly, in the Supply Chain Finance business, the transaction data are the trade confidentiality of each enterprise, and privacy protection becomes an essential and constantly evolving technology in order to alleviate the worries of enterprises when they put the transaction data on-chain. Some previous works such as literature\cite{ma2019privacy} studied the application of privacy protection mechanism of Hyperledger Fabric in Supply Chain Finance scenario. However, these privacy protection mechanisms were coarse-grained. Therefore, fine-grained privacy protection deserves a direction for in-depth research in the future, such as specific privacy protection mechanisms in different financing models, hierarchical subdivision of data access rights, different privacy data storage models, etc. This will meet the personalized privacy needs of different enterprises and different business scenarios, and also enrich the privacy protection technology system. Trustworthy data privacy protection technology will facilitate the development of Supply Chain Finance business and promote more companies to join the application of Blockchain technology.

\subsection{A comprehensive Blockchain-based platform for Supply Chain Finance.}
Combined with As a Service Platform and Financing Model Innovation in the previous section on hot-topics, a comprehensive Blockchain-based service platform is urgently needed and it may become a promising area to facilitate the application of Blockchain technology in the future, which allows users to not need to understand the esoteric theory of Blockchain but only need to know what functional modules the platform has and how to combine the modules to achieve various desired functions.In this comprehensive platform, Blockchain can be used as the infrastructure of the whole system, reflecting its function as a service platform. While on the other hand, various modules of Blockchain (such as privacy protection, smart contracts, consensus reaching and other sub-functions) can be combined into a certain financing mode of Supply Chain Finance, i.e. to demonstrate its function of enriching the financing mode.

Unlike uBaas\cite{lu2019ubaas}, which is mainly oriented to data management and smart contract applications, the comprehensive service platform proposed in this paper is oriented to the platform requirements and financing model requirements needed for Supply Chain Finance business. Meanwhile, distinguishing from Microsoft Azure\footnote{https://azure.microsoft.com/en-gb/solutions/Blockchain/. Last accessed: 2022-04-05}, IBM Hyperledger\footnote{https://www.ibm.com/Blockchain/platform/. Last accessed: 2022-04-05} and Amazon\footnote{https://aws.amazon.com/Blockchain/. Last accessed: 2022-04-05}, which lock users to a specific cloud or Blockchain platform, the comprehensive service platform proposed in this paper can meet the needs of enterprises to build Blockchain applications on their own private clouds. This comprehensive service platform only needs to no-chain the data of supply chain financing, and all their data are stored in enterprise’s private database. For the data of pre-financing evaluation and post-financing Supply Chain Finance business monitoring, it needs to be on-chained by the supply chain participants and participants to reach consensus through consensus mechanism. In order to avoid data falsification of a certain financing enterprise, it is necessary for its upstream and downstream enterprises or even the whole chain members to reach consensus on the data. This is similar to how a single node in a Blockchain cannot tamper with data because it requires most nodes to verify and reach consensus on the data, and it is difficult for a single financing company to falsify data. This platform does not depend on an exclusive company and platform, but is oriented to universal enterprises, and may be compatible with the original database of enterprises to meet the needs of Supply Chain Finance business while protecting the privacy and security of enterprise data. In other words, the new comprehensive Blockchain platform takes into account the needs of most enterprises' original Supply Chain Finance data or system convergence.

\subsection{Intelligent Risk Management for Supply Chain Finance.}

    \begin{figure}[h]
        \centering
        \vspace{-0.5cm}
        \setlength{\abovecaptionskip}{-0.3cm}
        \setlength{\belowcaptionskip}{-0.4cm}
        \includegraphics[width=\textwidth]{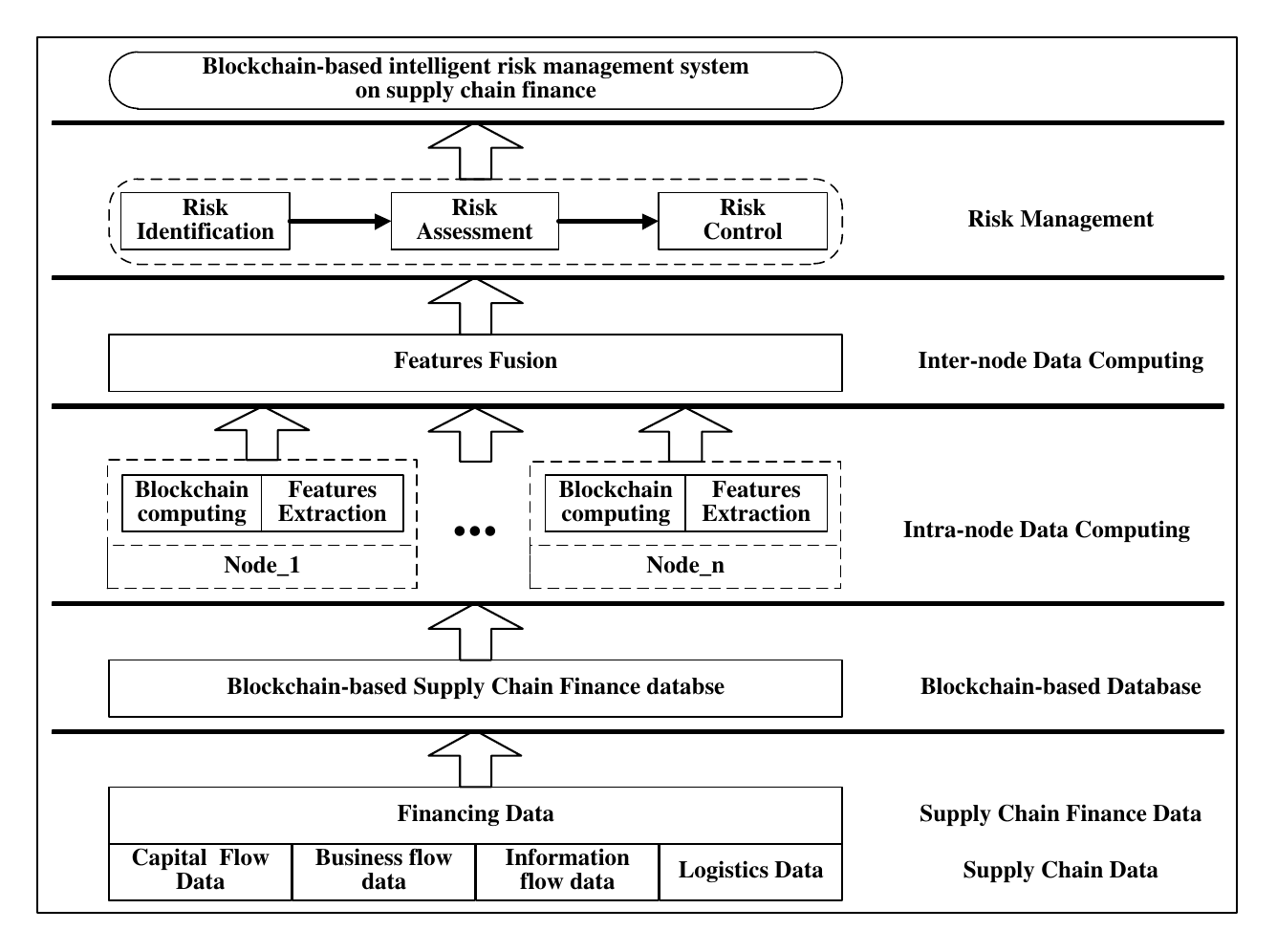}
        \caption{Intelligent Risk Management Conceptual Framework. In this framework, Blockchain computation and feature extraction can be done using computer techniques such as machine learning of artificial intelligence, data mining.}
        \label{fig:riskManagement}
    \end{figure}
    
Risk management is the identification, evaluation, and prioritization of risks followed by coordinated and economical application of resources to minimize, monitor, and control the probability or impact of unfortunate events or to maximize the realization of opportunities, and the main steps of risk management entail risk identification, risk assessment and risk control\footnote{https://en.wikipedia.org/wiki/Risk\_management. Last accessed: 2022-04-05}. That is, finding out what the risk is, measuring the magnitude of the risk and responding to it. Risk management is being prioritized by Supply Chain Finance and even all financing and credit operations. Benefiting from the development of big data, artificial intelligence and Blockchain technology, the research of risk management technology that is more intelligent and automated than today will become one of the most mainstream hot-topics of Blockchain-based Supply Chain Finance. In contrast to applying Blockchain to store data to alleviate information asymmetry, the advantage of Blockchain computing combining artificial intelligence and data mining techniques is that more information patterns can be learned progressively from Supply Chain Finance data to help risk control decision making. There have been some initial efforts in this direction. For example, literature\cite{liu2018using} and literature\cite{hu2020statistical} explored risk management problems in Supply Chain Finance using machine learning methods, while literature\cite{salman2018probabilistic} utilized probabilistic Blockchain to build efficient and distributed risk assessment and decision making applications. We believe that combining such work with Blockchain technology will result in Blockchain-based intelligent risk management technology. It is feasible if we use artificial intelligence techniques to extract data features directly from Blockchain data for risk identification and assessment. It is also reasonable to then utilize the risk assessment data to control the risk that exceeds the threshold value. In addition, it is also crucial to build a Blockchain-based Supply Chain Finance dataset and Blockchain computing for driving the development of intelligent risk control management technology. The Figure. {\ref{fig:riskManagement}} gives the conceptual diagram of our concept. The credit risk, collateral risk and operational risk of focus are all considered in intelligent risk management.

\subsection{Visualization of Blockchain-based Supply Chain Finance}
The visualization of Blockchain-based Supply Chain Finance will be a broadly applied research field, because visualization can display the invisible or not easily visible data patterns to people and help people understand the data status intuitively. Despite its active usage, Blockchain is a novel technology and its utilization in practice is still evolving and not well understood\cite{tovanich2019visualization}. To the best of our knowledge, in spite of the relatively few articles on Blockchain-based Supply Chain Finance visualization, a lot of work has been started on visualization of Blockchain technologies\cite{tovanich2019visualization, shahzad2021blockchain, putz2021hypersec} , such as Blockchain-based distributed data visualization\cite{shahzad2021blockchain}, research on Blockchain network attack detection visualization\cite{putz2021hypersec}. A highly relevant research work\cite{hao2020novel} was located by us, which proposed an approach combining Blockchain and visualization to mitigate the risk problem of food traceability. Hyperledger was utilized to store data and prevent data tampering. Visualization was used to assist people to visually analyze the causes of risk, such as heat maps to show non-conforming products, migration maps and force pointing maps to track products. Product traceability is to a certain extent based on the supply chain, which also inspires us: visualization technology is a good aid when supply chain financial data needs to be traced or risk analyzed. In addition, visualization can display the status of Supply Chain Finance data in real time, discover the risk points at the time of financing, monitor the business operation after financing, etc. Therefore, one important direction is to study Blockchain-based visualization of Supply Chain Finance.

\subsection{Cross-chain Research on Blockchain-based Supply Chain Finance.}
As the technology develops, the research of Blockchain-based Supply Chain Finance will develop from single chain to multiple chains. Because in practice, a certain enterprise may be included in multiple supply chains and Supply Chain Finance is based on supply chains, the best solution for financial institutions to evaluate a financing enterprise in a comprehensive and holistic way is to verify all supply chain operations in which the enterprise is involved. One of the problems that can be solved by cross-chain technology is to access and verify the status or events of other chains. Assuming that each supply chain is organized into a Blockchain network, Supply Chain Finance evaluates financing businesses from multiple supply chain data, which is equivalent to accessing and validating data across multiple Blockchains. The main ways of cross-chain are the aforementioned notary mechanism, side-chain/relay and hash lock. Recently, a novel cross-chain model in the form of oracle machine approach has emerged\cite{gao2020cross}, which was a framework that mainly entailed the source Blockchain, the data migration oracle machine and the destination Blockchain. The main steps were data migration oracle received user data migration request, triggered oracle smart contract deployed in Blockchain, then this oracle smart contract sent request to source Blockchain smart contract, the data sent by source Blockchain was encrypted and transmitted to data migration oracle. Then destination Blockchain smart contract sent request to oracle smart contract. The oracle smart contract retrieved the data from the data migration oracle and sent the data back to the destination Blockchain. This completed the cross-chain operation of migrating data from the source Blockchain to the destination Blockchain. In Blockchain-based Supply Chain Finance applications, financial institutions can acquire data from multiple Blockchains of Supply Chain Finance with the help of cross-chain technology to make more comprehensive and accurate asset and credit assessment of financing enterprises, which will be beneficial to reduce risks and facilitate the processing of financing business. Therefore, the research of cross-chain technology in the field of Blockchain Supply Chain Finance will be a promising direction.

\subsection{Blockchain Computing.}
Currently, Blockchain is considered as an improved version of distributed database because its main function is to provide a secure, reliable and decentralized shared data storage. However, it is limited to record only factual data, and the stored records are only valid 1 or invalid 0 two kinds of verification, i.e. Binary Validity limitation, and these recorded data are deterministic(such as A gave 10 coins to B) which can' t be probabilistic events(such as C only 90\% sure that A gave 10 coins to B).  However, a considerable number of events in life are probabilistic, which limits the scope of existing Blockchain applications. Therefore there is a need to explore a Blockchain model that can store deterministic data and events with probabilistic attributes. Inspired by two talks, "Extending Blockchains with AI for Risk Management\footnote{https://www.cse.wustl.edu/\textasciitilde jain/talks/pbc\_iics.htm. Last accessed: 2022-04-01} "and "Extending Blockchains for Risk Management and Decision Making\footnote{https://www.cse.wustl.edu/\textasciitilde jain/talks/pbc\_ibf.htm. Last accessed: 2022-04-01} ", given by Professor Raj Jain\cite{salman2019reputation, salman2018probabilistic}, we constructively propose a new concept: Blockchain computing. Blockchain computing is defined as a type of Blockchain that combines data storage and computing functions. It consists of computation within a single block, called intra-Blockchain computing, and computation between multiple blocks, called inter-Blockchain computing. The computation method can be a statistical algorithm, data mining algorithm or artificial intelligence algorithm embedded in a block or Blockchain. For intra-Blockchain computing, the expression can take the form of a block summary structure within a block, which is the result of the computation on transactions or events within a block, expressed as a feature vector or probability score, etc. For inter-Blockchain computing, the exhibited form can be the result of a block's computation on multiple previous blocks. Whether it is intra-Blockchain computing or inter-Blockchain computing, a new block is endorsed by consensus of all nodes before it is integrated into the Blockchain, which is consistent with the traditional block's on-chain model. Blockchain computing can achieve extending stored deterministic data to probabilistic events, extending storage functions to computing functions, and extending data to information knowledge. Such a Blockchain can be perfectly fused with various popular technologies such as artificial intelligence, big data mining, and cloud computing.

\subsection{Extending smart contract for Blockchain-based Supply Chain Finance.}
As introduced in the application hot-topics, smart contract can achieve the pre-defined action triggered when the trigger condition reaches the threshold value, this automation function can help Supply Chain Finance reduce the errors caused by human operation. Smart contract was widely applied to risk management, credit management, data management, platform construction and financing model transformation in Supply Chain Finance. However, these applications were preliminary. It is worth exploring how to use smart contract in a broader and deeper way. For example, the work\cite{chen2020blockchain} partially automated the workflow of Supply Chain Finance through smart contract. The business system of Supply Chain Finance is enormous, with diverse financing models, complex financing operations, and intertwined financing supply chains, which bring great challenges to the application of smart contract. In addition to expanding the scope of application, smart contract could be combined with current advanced computer technology, such as machine learning\cite{eshghie2021dynamic} and data mining\cite{jung2019data}, to calculate the more suitable threshold parameters. Alternatively, adaptive multi-functional smart contract can be studied to meet the Supply Chain Finance needs of different scenarios. The incorporation of smart contract is a cross-era symbol of Blockchain technology, which expands the application scope of Blockchain technology and closes the distance between Blockchain theory and practical application. We deeply believe that expanding the research of smart contract has a very positive significance both for the development of Blockchain technology and for promoting the automation and intelligence of Supply Chain Finance.

\subsection{Other Promising Research directions on Blockchain-based Supply Chain Finance.}
In addition to the data foundation, platform construction, application scenarios and technical expansion we have already mentioned, there are many interesting research directions on Blockchain-based Supply Chain Finance. Such as government auditing, as a social regulator needs to inspect financial institutions and financing companies about Supply Chain Finance business, Blockchain can provide them with anti-falsification and anti-tampering data records, so how to help regulators conduct auditing efficiently and effectively is also a direction worth researching. Another example is how to take advantage of the distributed nodes of Blockchain network to help financial institutions complete the evaluation and lending when financing in a federated computing mode, such as federal learning\cite{hou2021systematic}, ensemble learning\cite{jatoth2021improved, yu2019decentralized}, etc. Also how to construct Blockchain machine\cite{javaid2021blockchain} to help Supply Chain Finance data on-chain and enterprise-financial institution environment deployment, and how to combine metaverse, AI and Blockchain technologies\cite{ynag2022fusing, gadekallu2022blockchain} to open new horizons for Supply Chain Finance research. In spite of the fact that these technologies combined with Blockchain have not yet been fully applied to Supply Chain Finance, these works have given an indication of the possibilities. In any case, many meaningful research directions are worth exploring to improve the Blockchain-based Supply Chain Finance system.

\section{conclusions} \label{title:conclusions}
Supply Chain Finance has a significant impact on the innovation of financing solutions, the improvement of supply chain competitiveness and the application of advanced computer technologies in practice. The application of Blockchain, emerging computer technology, to Supply Chain Finance can improve its financing system, promote its intelligence and automation, and provide more convenient financing and cost optimization for the supply chain. In this survey, we briefly review the basic knowledge and current research progress of Supply Chain Finance and Blockchain, respectively. Then, some hot applications brought by the combination of Blockchain and supply chain are summarized: risk management, service platform function, financing model innovation, credit management, data management and smart contract application. Finally, Blockchain-based Supply Chain Finance research directions such as building Blockchain Supply Chain Finance dataset, data privacy protection, comprehensive Blockchain platform, intelligent risk management, visualization, cross-chain, Blockchain computing and expanding smart contract are proposed. These promising directions need further research.

\bibliographystyle{unsrt}  
\bibliography{references}

\end{document}